\documentclass[usenatbib,usegraphicx]{mn2e}
\usepackage{comment}
\usepackage{amsfonts}
\usepackage{amsmath}

\title[Breaking the age-metallicity degeneracy using $UV$ photometry]
    {Better age estimation using UV-optical colours:
breaking the age-metallicity degeneracy}

\author[S.~Kaviraj et al.]
{S. Kaviraj$^{1}$\thanks{E-mail: skaviraj@astro.ox.ac.uk}, S. -C.
Rey$^{2,3,4}$, R. M. Rich$^{5}$, S. -J. Yoon$^{3}$ and S. K. Yi$^{3}$\\
$^{1}$ Department of Physics, University of Oxford, Keble Road,
Oxford OX1 3RH, UK\\
$^{2}$ Department of Astronomy and Space Science, Chungnam National University, Daejeon 305-764, Korea\\
$^{3}$ Yonsei University, Center for Space Astrophysics, Seoul
120­749, Korea\\
$^{4}$ California Institute of Technology, 1200 E. California
Blvd. Pasadena, CA 91125, USA\\
$^{5}$ Department of Physics and Astronomy, UCLA, 430 Portola
Plaza, Box 951547, Los Angeles, CA 90095-1547}

\begin{document}

\vspace{-0.1in}

\date{19 November 2006}

\pagerange{\pageref{firstpage}--\pageref{lastpage}} \pubyear{2003}

\maketitle

\label{firstpage}


\begin{abstract}
We demonstrate that the combination of GALEX $UV$ photometry in the $FUV$ ($\sim$ 1530
angstroms) and $NUV$ ($\sim$ 2310 angstroms) passbands with optical
photometry in the standard $U,B,V,R,I$ filters can
efficiently break the age-metallicity degeneracy. We estimate well-constrained
ages, metallicities and their associated errors for 42 GCs in M31, and
show that the full set of $FUV,NUV,U,B,V,R,I$ photometry produces
age estimates that are $\sim$ 90 percent more constrained and
metallicity estimates that are $\sim$ 60 percent more constrained than
those produced by using optical filters alone. The quality of the age constraints is
comparable or marginally better than those achieved using a large number of spectroscopic
indices.
\end{abstract}




\vspace{-0.3in}

\section{Introduction}
Broad-band photometry has traditionally played a key role in the study of a variety
of astrophysical objects. Often, it is of interest to
decipher the formation histories of various systems, e.g. globular
clusters (GCs), by age-dating them. The vast majority
of studies have so far been restricted to \emph{optical}
photometry e.g. \citep{Yi2004,Peng2004}. However, the use of
optical photometry has been plagued by the well-known
\emph{age-metallicity degeneracy} (AMD) whereby young, metal-rich stellar populations produce optical
colours which are indistinguishable from those due to old metal-poor
populations \citep{Worthey1994}. It is therefore difficult to simultaneously pin
down the age and metallicity of an object using optical colours
alone. While combinations of metallicity-sensitive and age-sensitive
colours, e.g. $U-B$ versus $B-V$  \citep{Yi2004}, can produce rough
estimates of age and metallicity, the errors on these estimates can be
extremely large. For example, the age uncertainty could be as large as
the age of the universe \citep[e.g.][]{Yi2004}!

Ages can be better studied, and the AMD alleviated
spectroscopically, by combining a sufficient number of age and
metallicity-sensitive line indices \citep[see
e.g.][]{Trager2000,Proctor2002,Caldwell2003}. For
example, \citet{Trager2000} use the $H_{\beta}$, Mg $b$ and
$<$Fe$>$ line strengths to estimate well-constrained
SSP-equivalent ages and metallicities for a sample of local
elliptical galaxies. Similarly, \citet{Proctor2002} use a
combination of 24 line indices to break the AMD and produce
SSP-equivalent parameters with small uncertainties for the central
regions of 32 systems taken from both the Virgo cluster and the field. However, such methods
rely on accurate measurements of a sufficiently large ensemble of
indices for each object of interest.

Rest-frame $UV$ flux, shortward of $\sim$ 3000 angstroms, provides
an attractive alternative. From a theoretical point of view, it
has already been suggested \citep[e.g.][]{Yi2003} that the $UV$
spectral ranges may provide a better handle on the ages of objects
and thus reduce the AMD. It is the aim of this Letter to
\emph{quantify} the usefulness of the $UV$ in
breaking the AMD and providing robust age estimation, especially
in the context of simple populations like GCs. If the addition of
$UV$ photometry were to provide comparable constraints in the
age-metallicity parameter space as, for example, large numbers of
spectral line indices, then the advent of large-scale $UV$
photometry from the GALEX mission (Martin et al. 2005) could provide an accessible
means of age-dating a wide variety of nearby objects using
photometry alone. For example, in its Medium Imaging Survey (MIS)
mode, which reaches a depth of 22.6 and 22.7 AB in $FUV$ and $NUV$
respectively, GALEX detects more than 90 percent of
bright ($r<16.8$), nearby ($0<z<0.11$) early-type galaxies observed by
the SDSS (see Figure 1 in Kaviraj et al. 2006). Given the large-scale
nature of the GALEX dataset and the excellent overlap with existing
optical surveys, it is therefore possible to incorporate $UV$
photometry into age-dating analyses of stellar populations. 

In this study we combine GALEX broad-band $UV$ photometry in two
passbands ($NUV$: $\sim$ 2310 angstroms and $FUV$: $\sim$ 1530
angstroms) with optical photometry in the standard $U,B,V,R,I$
filters to simultaneously estimate the ages and metallicities of
GCs in M31. The GCs are drawn from the catalogue of Rey et al.
(2005) who have combined GALEX imaging of the M31 system with the
optical catalogue of \citet{Barmby2000}. The use of GCs is
motivated by the fact that their star formation histories (SFHs)
can be adequately parametrised by simple stellar populations
(SSPs) of a given age and metallicity alone. They are therefore
ideal test objects to demonstrate the usefulness of $UV$
photometry in constraining the age-metallicity parameter space.
Although we derive both age and metallicity, our emphasis is on
the quality of the \emph{age estimation}, as this is of primary
importance in the determination of formation histories. Note that,
while the stellar models used here (described below) assume solar abundance
ratios, GC age estimates derived using \emph{optical} colours tend to
be lower by $\sim8$ percent if alpha-enhanced $([\alpha/Fe]=0.3)$ ratios are
considered \citep{Kim2002}.

In the following sections, we compare the quality of the
constraints on age and metallicity with and without $UV$
photometry, investigate how the constraints change as various
filters are excluded from the parameter estimation and compare our
purely photometric analysis to previous studies that have used
spectroscopic indicators to estimate SSP-equivalent ages and
metallicities for a variety of systems.


\vspace{-0.1in}

\section{Parameter estimation}
We begin by outlining the method used for parameter
estimation. For a vector $\textbf{X}$ denoting parameters in the model and a
vector $D$ denoting the measured observables (data),

\begin{equation}
\textnormal{prob}(\textbf{X}|D) \propto \textnormal{prob}(D|\textbf{X}) \times \textnormal{prob}(\textbf{X}),
\end{equation}

where $\textnormal{prob}(\textbf{X}|D)$ is the probability of the model given the data
(which is the object of interest), $\textnormal{prob}(D|\textbf{X})$ is the probability of
the data given the model and $\textnormal{prob}(\textbf{X})$ is the prior probability
distribution of the model parameters. Assuming a uniform prior in our
model parameters and gaussian errors gives





\begin{equation}
\textnormal{prob}(D|\textbf{X}) \propto \exp(-\chi^2/2),
\end{equation}

where $\exp(-\chi^2/2)$ is the likelihood function and $\chi^2$ is
defined in the standard way. $\textnormal{prob}(\textbf{X}|D)$ is a \emph{joint} PDF, dependent
on all the model parameters. To isolate the effect of a single
parameter $X_1$ in, for example, a two parameter model
($\textnormal{prob}(\textbf{X}|D) \equiv
\textnormal{prob}(X_1,X_2|D)$) we can integrate out the effect of
$X_2$ to obtain the \emph{marginalised} PDF for $X_1$:




\begin{equation}
\textnormal{prob}(X_1|D)=\int^{\infty}_0 \textnormal{prob}(X_1,X_2|D)dX_2.
\end{equation}

In our analysis, the parameters $\textbf{X}$ in the model are the
age ($t$) and metallicity ($Z$) of the SSP in question, and the
observables are colours constructed from $FUV,NUV,U,B,V,R,I$
photometry. We choose a uniform prior in $t$, in the range 0 to 14
Gyrs. We set the maximum allowed age to 14 Gyrs using the WMAP
estimate for the age of the universe (13.7 Gyrs, Spergel et
al. 2003) - we note that structure formation probably commences
$\sim$ 0.2-0.5 Gyrs after the creation of the Universe. For the
metallicity $Z$, we use a uniform prior in the range -2.3 to 0.36
dex, which are essentially the metallicity limits of the stellar
models. For each GC analysed, we marginalise the age and
metallicity parameters, take the \emph{peak} ($x_0$) of the
marginal PDF $P(x)$ as the best estimate and \emph{define}
one-sigma limits $x_+$ and $x_-$ as $\int_{x_0}^{x_{+}} P(x)dx = 0.34
\int_{x_0}^\infty P(x)dx$ and $\int_{x_{-}}^{x_0} P(x)dx = 0.34
\int_{-\infty}^{x_0} P(x)dx$ respectively.

\begin{figure}
\begin{center}
\includegraphics[width=3.5in]{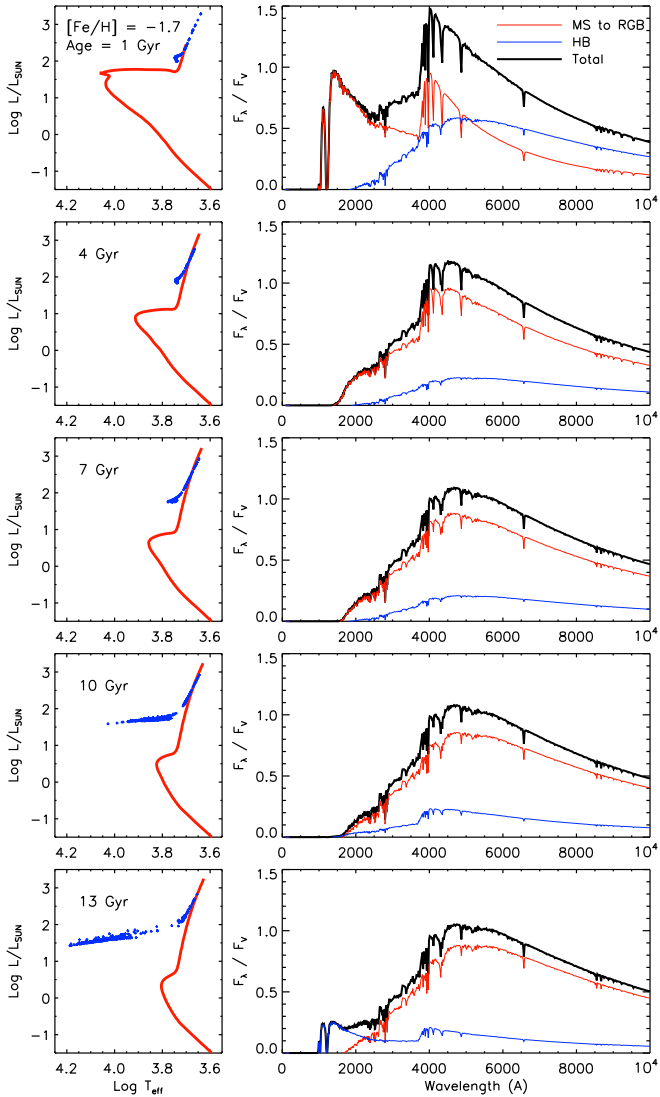}
\caption{$UV$ flux from old populations due to the onset
  of the horizontal branch population (top to bottom). Intermediate age
populations ($\sim$ 4-7 Gyrs old) output negligible amounts of $UV$
  flux. The contributions from the MS-RGB and HB phases are shown in
  red and blue respectively. In all panels [Fe/H] = -1.7 dex. The HB
  develops at higher (lower) ages for [Fe/H] greater (lower) than -1.7
  dex. For example, for [Fe/H]$\sim$-1.2 and [Fe/H]$\sim$-2.1 dex, the
  HB develops at 12 and 8 Gyrs respectively. Note that this implies
  that metal-rich GCs will have weaker age
  constraints due to the later rise of the HB.} 
\label{fig:old_uv_flux}
\end{center}
\end{figure}

\begin{figure}
\begin{center}
\includegraphics[width=3.5in]{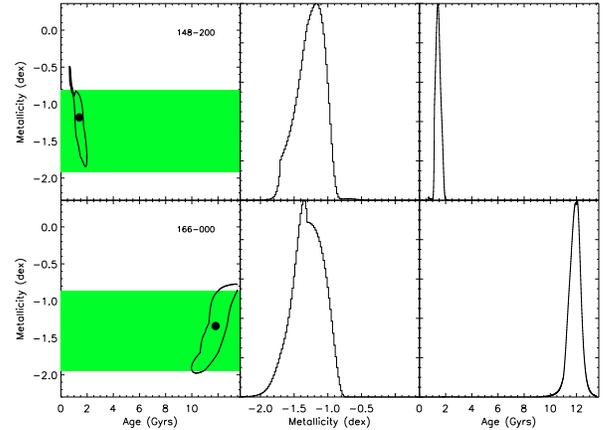}
\caption{Age and metallicity constraints from $UV$ and
optical photometry for two GCs in our sample. LEFT COLUMN:
Best-fit position and one-sigma contour. MIDDLE COLUMN:
Marginalised metallicity PDF (normalised). RIGHT COLUMN:
Marginalised age PDF (normalised). Top row shows a young cluster
(148-200) and bottom row shows an old cluster (166-000). The
  green region shows a spectroscopic metallicity measurement taken
  from \citet{Barmby2000} or \citet{Perrett2003}.}
\label{fig:all_colour_examples}
\end{center}
\end{figure}

The \emph{uncertainty} in the parameter is then ($x_{+} -x_{-}$). Note
that we do not consider correlated errors in the two quantities.


\vspace{-0.1in}

\section{Age and metallicity constraints}
Young stellar populations ($<1$ Gyr old) emit a substantial
proportion of their flux in the ultraviolet, making $UV$
photometry a good indicator of their presence. However, $UV$ flux
can also be generated by old, evolved stages of stellar evolution.
Due to their longevity, hot, core Helium-burning horizontal-branch
(HB) stars and their progeny are thought to be the dominant $UV$
sources in old stellar populations, such as GCs and elliptical
galaxies. While metallicity is the dominant parameter controlling
the HB morphology of Galactic halo-like stellar populations, the
second parameter governing the properties of the HB has been the subject
of much debate \citep[e.g.][]{Sarajedini1997,Bellazzini2001},
concensus favouring age differences between clusters as the major
second parameter affecting HB morphology \citep[e.g.][]{Lee1994a}. 

The stellar models used in this study employ detailed prescriptions
for the systematic variation of HB morphology with metallicity and
age, and are well-calibrated to Galactic GC populations and the
integrated $UV$-optical photometry of local elliptical galaxies
\citep[see][and references therein]{Yi2003}. Main-sequence to Red
Giant Branch isochrones employed in these models are described in
\citet{Yi2001} and the HB stellar tracks are described in
\citet{Yi1997}. The flux library employed is from
\citet{Lejeune1998}. 

In Figure \ref{fig:old_uv_flux} we illustrate the onset of $UV$ flux in the
models with the appearance of horizontal branch stars,
as the stellar population evolves through intermediate ages of 4
and 7 Gyrs to an age comparable to stars in the Galactic halo
($\sim$ 13 Gyrs). The age estimation
procedure essentially fits the observed shape of the $UV$-optical
continuum to the models. In this context we note that, although
the shape of the $UV$ spectrum from a young and old population
can appear similar, the slope of their optical continua are very
different, which allows them to be differentiated easily. Thus,
with access to a reasonably large set of $UV$ and optical filters,
age determinations of stellar populations such as GCs can be
performed with reasonably good accuracy, as we demonstrate below.

The purpose of this Letter is not to perform a rigorous study of
the entire M31 GC system but to demonstrate the usefulness of $UV$
photometry in reducing the AMD and present a method that exploits
this property. We therefore extract parts of Rey et al. (2005)'s
catalogue that best satisfy this aim - GCs with good detections in
GALEX, which have the full set of optical $U,B,V,R,I$ photometry
and lie in regions of relatively low Galactic reddening
($E_{B-V}<0.2$). We find 42 objects matching this description in
the catalogue.

\begin{figure}
\begin{center}
\includegraphics[width=3.5in]{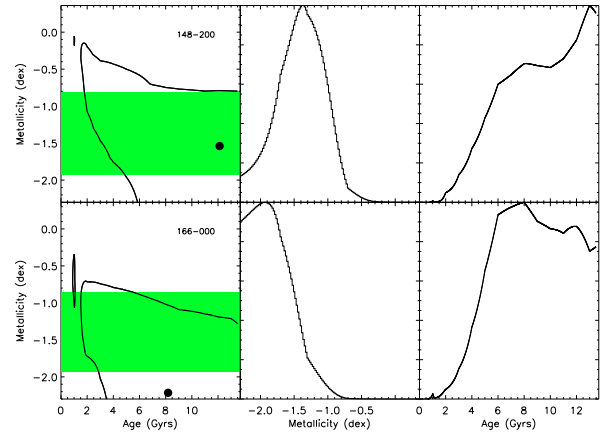}
\caption{Same as Figure \ref{fig:all_colour_examples} but using optical
photometry only.}
\label{fig:optical_examples}
\end{center}
\end{figure}

We begin by investigating the age and metallicity constraints
achieved using seven broad-band colours: $FUV-V$, $NUV-V$,
$FUV-NUV$, $U-B$, $B-V$, $V-R$, $V-I$ (Figure
\ref{fig:all_colour_examples}) and compare them to the scenario
where the $UV$ colours are absent (Figure
\ref{fig:optical_examples}). The large AMD due to using only
optical passbands is clearly visible in Figure
\ref{fig:optical_examples}. The effect on the age estimation is
severe - since the parameter space inside the contour is
indistinguishable in terms of the goodness of fit (the
\emph{best-fit} value has very limited significance), the age of
either GC can be anywhere between 2 and 13 Gyrs. The age
constraint is far superior when $UV$ photometry is added. The
young cluster (148-200) now has a well-constrained age of 0.5-2
Gyrs and the old cluster has an age of 10-13 Gyrs. The photometry
suggests that both clusters are metal-poor ([Fe/H] $< -0.5$), in
agreement with the spectroscopically determined metallicity (green
region).

\begin{figure}
\begin{center}
\includegraphics[width=3.5in]{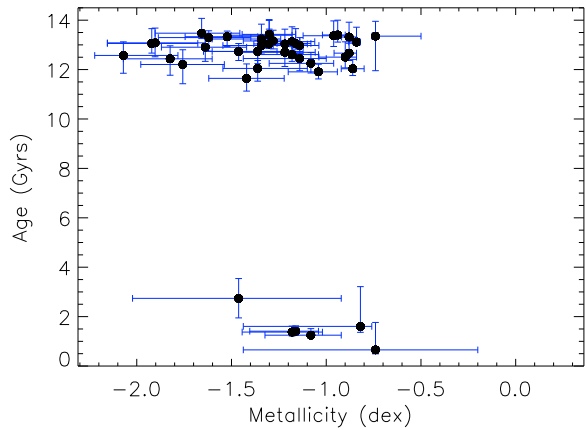}
\caption{Age and metallicity constraints from the full set of $UV$
and optical photometry from our 42 GCs.}
\label{fig:age_metallicity_constraints}
\end{center}
\end{figure}

In Figure \ref{fig:age_metallicity_constraints} we summarise the
age and metallicity constraints from the full set of $UV$ and
optical photometry for the 42 GCs in our sample. The age
constraints are robust, with the full extent of the age
uncertainties lying between 0.1 and 3 Gyrs. The metallicity
constraint is not as robust as the age uncertainties, which is
expected since the $UV$ is a better indicator of age than
metallicity. We find that, although the photometric metallicity
constraint overlaps with its spectroscopic equivalent in all
cases, the spectroscopic constraint is invariably more robust.

\begin{figure}
\begin{center}
\includegraphics[width=3.5in]{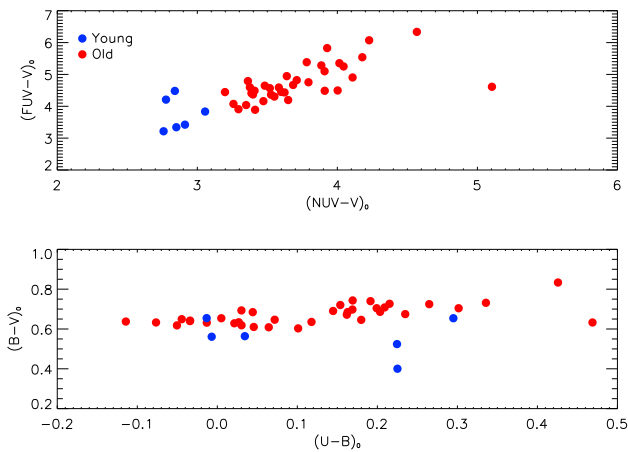}
\caption{TOP PANEL: Comparison of the best-fit ages of GCs to
their $(NUV-V)_0$ and $(FUV-V)_0$ colours. BOTTOM PANEL:
Comparison of the best-fit ages of GCs to their $(U-B)$ and $(B-V)$
colours.} \label{fig:colour_separation}
\end{center}
\end{figure}

In Figure \ref{fig:colour_separation} we compare the best-fit ages
of each GC to their photometric colours. Contrary to the optical
scenario, we can use the best-fit ages in this case \emph{because
the age uncertainties are small}. We find that GCs with
well-constrained young ages tend to be bluer in both the
$(NUV-V)_0$ and $(FUV-V)_0$ colours. In contrast, the comparison
to optical colours is more ambiguous - while there is an
indication that the young GCs are bluer in $(B-V)_0$, there is no
clear separation between the young and old populations, as a result of
the AMD. 

\begin{table*}
\begin{center}
\caption{Comparison of the quality of age and metallicity
constraints for different filter sets. $f_t$ ($f_Z$) is the ratio of the
age (metallicity) uncertainty using the filters in a given row to the age
uncertainty using optical filters only (row 7). The filter
sets are listed in order of the quality of the \emph{age
constraint} i.e. the first row indicates the filter set with the
\emph{best} age constraint.}

\begin{tabular}{cccccccll}

    & $FUV$ & $NUV$ & $U$ & $B$ & $V$ & $R,I$ & $f_t$ & $f_Z$\\ \hline \hline

    1 & \checkmark & \checkmark & \checkmark & \checkmark 
    & \checkmark & \checkmark & 0.12 & 0.31\\

    2 & \checkmark & \checkmark & \checkmark & \checkmark
    & \checkmark & $\times$   & 0.23 & 0.37\\

    3 & \checkmark & \checkmark & $\times$ & \checkmark 
    & \checkmark & \checkmark & 0.26 & 0.38\\

    4 & \checkmark & \checkmark & $\times$ & \checkmark &
    \checkmark & $\times$ & 0.26 & 0.39\\

    5 & \checkmark & $\times$ & \checkmark & \checkmark &
    \checkmark & \checkmark & 0.39 & 0.47\\

    6 & $\times$ & \checkmark & \checkmark & \checkmark &
    \checkmark & \checkmark & 0.47 & 0.51\\

    7 & $\times$ & $\times$ & \checkmark & \checkmark &
    \checkmark & \checkmark & 1.00 & 1.00

\end{tabular}
\end{center}
\label{tab:constraint_comparisons}
\end{table*}

In practice, a full set of $FUV,NUV,U,B,V,R,I$ photometry may not
always be available. For example, good quality photometry is far
more difficult to achieve in the $U$ band than in longer optical
wavelengths. It is therefore instructive to compare the quality of
the age and metallicity constraints when a subset of these
passbands is excluded from the parameter estimation. We quantify
this comparison in Table 1. Since using only optical filters
always gives the largest uncertainties, we compute the ratio of
the uncertainty from using a particular set of filters to the
uncertainty when only optical filters are used. We show the
average values of these ratios, for the age constraints ($f_t$ -
column 9) and for the metallicity constraints ($f_Z$ - column 10).
A smaller value thus indicates a \emph{better} constraint and
smaller uncertainty in the estimation of age or metallicity. The
filter sets are listed in order of the quality of the \emph{age
constraint} i.e. the first row indicates the filter set with the
\emph{best} age constraint. The constraints are virtually identical
as long as either the $R$ or $I$ filters are present - we have
therefore put them in the same column in Table 1.

Looking first at constraints on the age estimates, we find that, for the full set of
$FUV,NUV,U,B,V,R,I$ filters (row 1), the size of the marginalised age
errors are $\sim10$ percent of the optical-only case ($f_t = 0.12$) i.e. the age estimates are $\sim$ 90 percent better
constrained than those from using optical filters alone (row 7). Removing the longer
wavelength $R,I$ bands (row 2) or the shorter wavelength $U$ band
(row 3) slightly weakens the age constraint compared to the full
set of filters. If only one $UV$ filter is available, it is
preferable to combine the $FUV$ passband with optical photometry
(row 5) than the $NUV$ passband (row 6).

Similarly, the comparison of metallicity estimates (values of $f_Z$ in Table
1) indicates that adding $UV$ photometry to the optical
$U,B,V,R,I$ filters (or a subset of them) improves the metallicity
estimates by approximately $\sim$ 60 percent compared to the
optical-only case. $f_Z$ refers to the underlying
metallicity $Z$, converted from the [Fe/H] value using Table 2 of
\citet{Yi2001}.


\vspace{-0.1in}

\section{Comparison with spectroscopic methods}
We now compare the quality of our photometric age and metallicity
constraints with three studies which derive these parameters using
spectroscopic line indices. The \citet{Trager2000} study uses the
$H_{\beta}$, Mg $b$, Fe5270 and Fe5335 indices to obtain
tightly-constrained SSP ages for a sample of local elliptical
galaxies, with age uncertainties of $\sim 0.5-1.5$ Gyrs for young
objects and $\sim 2-3$ Gyrs for old and intermediate age objects
(see their Table 6A). In a similar work, \citet{Proctor2002}
combine line strength measurements of 24 age and metallicity
sensitive indices for the central regions of 32 galaxies in
both the Virgo cluster and the field. Their age uncertainties are in the range $\sim 1-3$
Gyrs (see their Table 10 or Figure 12). In a more recent work,
\citet{Beasley2004} measure 24 indices in the Lick system
\citep{Worthey1997,Trager1998} for a sample of 30 globular
clusters in M31. The uncertainties in their age-estimates are
$\sim$ 0.2-1.25 Gyrs (see their Figure 12). Looking at Figure
\ref{fig:age_metallicity_constraints} we see that the
uncertainties in our photometric age-determinations are $\sim
0.1-2$ Gyrs for the young clusters and $\sim 0.5-3$ Gyrs for the
old clusters. We therefore find that our photometric analysis
produces comparable age constraints to the spectroscopic studies.

However, our metallicity constraints are arguably worse than those
in the spectroscopic studies mentioned above. While we do not
resolve the metallicity to better than $\pm 0.3$ to 0.5 dex for
any cluster, spectroscopic constraints on metallicity are
generally tighter, of the order of $\sim 0.1$ dex. This is
expected, not only since the $UV$ passbands are better indicators
of age than of metallicity but because the combination of multiple
metallicity-sensitive indices in these studies results in better
metallicity constraints than can be achieved with photometry
alone.


\vspace{-0.28in}

\section{Conclusions}
In this study we have demonstrated that the combination of $FUV$
($\sim$ 1530 angstroms) and $NUV$ ($\sim$ 2310 angstroms)
photometry with optical observations in the standard $U,B,V,R,I$
filters can efficiently break the AMD caused by using optical
filters alone. We have estimated ages, metallicities and their
associated uncertainties for a sample of 42 GCs in M31 for which
we have the full set of $UV$ and optical photometry. We have found
that using the full set of $FUV,NUV,U,B,V,R,I$ filters produces
age estimates that are $\sim$ 90 percent more constrained and
metallicity estimates that are $\sim$ 60 percent more constrained
than those produced by using optical filters alone. The quality
of the age constraints is comparable to those achieved using a
large number of spectroscopic line strength indices. We find that
if only one $UV$ passband is available, then combining the $FUV$
rather than the $NUV$ with optical photometry is preferable in
terms of constraining ages. Finally, while there is overlap
between our photometrically determined metallicity ranges and the
corresponding constraints from spectroscopic studies for GCs in
our sample, the photometric metallicity constraints are generally
weaker than their spectroscopic counterparts. 

We note that, since the $UV$ is produced by either young ($<2$ Gyr
old) or evolved ($>10$ Gyr old) stars, it is insensitive to
intermediate-age populations, leading to a lack of intermediate-age
GCs in Figure 4. To conclude, we suggest that broad-band $UV$
photometry from the GALEX mission (which, at completion, will survey
the entire local universe) could be used as an effective age indicator
to break the AMD, and as a useful proxy for spectroscopic methods in
constraining the age-metallicity parameter space.


\nocite{Spergel2003}
\nocite{Rey2005}
\nocite{Martin2005}
\nocite{Kaviraj2006}


\vspace{-0.1in}

\section{Acknowledgements}
We thank Andr\'es Jord\'an, Roger Davies, Rachel Somerville and
Claudia Maraston for many useful discussions. We are grateful to the
referee, Anne Sansom, for an insightful review that greatly improved
the quality of the manuscript. SK
acknowledges a PPARC DPhil scholarship, a Leverhulme Early-Career
Fellowship and Research Fellowships from Worcester College, Oxford and
the BIPAC institute at Oxford. SCR acknowledges support through the
Korea Research Foundation Grant, funded by the Korean Government
(MOEHRD) (KRF-2005-202-C00158). SJY and SKY acknowledge support through
Grant No. R01-2006-000-10716-0 from the Basic Research Program
of the Korea Science \& Engineering Foundation.


\vspace{-0.1in}

\bibliographystyle{mn2e}
\bibliography{references}


\end{document}